\documentclass[a4paper]{jpconf}
\usepackage{graphicx}
\usepackage{caption}
\usepackage{subcaption}
\usepackage{xcolor}
\usepackage{acronym}
\usepackage{amsmath}
\usepackage{amssymb}
\usepackage{cleveref}
\usepackage{acronym}
\usepackage{xspace}
\usepackage{booktabs}

\usepackage[
    hyperref=true,
    url=false,
    isbn=false,
    backref=false,
    style=numeric,
    block=none,
    doi=true,
    eprint=false,
    sorting=none
]{biblatex}
\bibliography{bibliography}

\acrodef{ML}{machine learning}
\acrodef{PF}{particle-flow}
\acrodef{JetMET}{jets and missing energy}
\acrodef{DQM}{data quality monitoring}
\acrodef{GNN}{graph neural network}
\acrodef{LHC}{CERN Large Hadron Collider}
\acrodef{MLPF}{machine-learned particle-flow}
\acrodef{GCN}{graph convolutional network}
\acrodef{LSH}{locality sensitive hashing}
\acrodef{GPU}{graphics processing unit}
\acrodef{ECAL}{electromagnetic calorimeter}
\acrodef{HCAL}{hadron calorimeter}
\acrodef{PU}{pileup}
\acrodef{IPU}{intelligence processing unit}
\acrodef{MC}{Monte Carlo simulation}
\acrodef{CMSSW}{CMS software framework}
\acrodef{MET}{missing transverse energy}
\acrodef{GSF}{Gaussian sum filter}
\acrodef{DT}{drift tube}
\acrodef{CSC}{cathode strip chamber}
\acrodef{MC}{Monte Carlo}
\acrodef{HF}{hadron forward}
\acrodef{HFEM}{HF electromagnetic energy}
\acrodef{HFHAD}{HF hadronic energy}
\acrodef{BREM}{Bremsstrahlung}
\acrodef{KF}{Kalman filter}
\acrodef{LHC}{Large Hadron Collider}

\newcommand{\qcd}{QCD\xspace}
\newcommand{\ttbar}{\ensuremath{\mathrm{t}\overline{\mathrm{t}}}\xspace}
\newcommand{\PZ}{\ensuremath{\mathrm{Z}}\xspace}

\newcommand{\unit}[1]{\ensuremath{\text{\,#1}}\xspace}
\newcommand{\PYTHIA} {{\textsc{Pythia}}\xspace}
\newcommand{\DELPHES} {{\textsc{delphes}}\xspace}
\newcommand{\TENSORFLOW} {{\textsc{TensorFlow}}\xspace}
\newcommand{\ONNX} {{\textsc{ONNX}}\xspace}
\newcommand{\ONNXRUNTIME} {{\textsc{onnxruntime}}\xspace}

\newcommand{\ptmomentum}{\ensuremath{p_{\mathrm{T}}}\xspace}
\newcommand{\pseudorapidity}{\ensuremath{\eta}\xspace}
\newcommand{\akfourchs}[1]{AK4-CHS\xspace}
\newcommand{\akfourpuppi}[1]{AK4-PUPPI\xspace}
\newcommand{\GeV}{\ensuremath{\,\text{Ge\hspace{-.08em}V}}\xspace}

\begin{document}
\title{Machine Learning for Particle Flow Reconstruction at CMS}

\author{Joosep Pata$^{1,*}$, Javier Duarte$^2$, Farouk Mokhtar$^2$, Eric Wulff$^3$, Jieun Yoo$^4$, Jean-Roch Vlimant$^5$, Maurizio Pierini$^3$, Maria Girone$^3$\\
{\normalfont (on behalf of the CMS Collaboration)}}

\address{$^1$NICPB, R\"{a}vala pst 10, 10143 Tallinn, Estonia}
\address{$^2$University of California San Diego, La Jolla, CA 92093, USA}
\address{$^3$European Center for Nuclear Research (CERN), CH 1211, Geneva 23, Switzerland}
\address{$^4$University of Illinois Chicago, Chicago, IL 60605, USA}
\address{$^5$California Institute of Technology, Pasadena, CA 91125, USA}

\ead{$^*$joosep.pata@cern.ch}

\begin{abstract}
We provide details on the implementation of a machine-learning based particle flow algorithm for CMS.
The standard particle flow algorithm reconstructs stable particles based on calorimeter clusters and tracks to provide a global event reconstruction that exploits the combined information of multiple detector subsystems, leading to strong improvements for quantities such as jets and missing transverse energy.
We have studied a possible evolution of particle flow towards heterogeneous computing platforms such as GPUs using a graph neural network.
The machine-learned PF model reconstructs particle candidates based on the full list of tracks and calorimeter clusters in the event.
For validation, we determine the physics performance directly in the CMS software framework when the proposed algorithm is interfaced with the offline reconstruction of jets and missing transverse energy. 
We also report the computational performance of the algorithm, which scales approximately linearly in runtime and memory usage with the input size.
\end{abstract}

 \section{Introduction\label{sec:intro}}
Reconstruction algorithms at general-purpose high-energy particle detectors aim to provide a holistic, well-calibrated physics interpretation of the collision event. 
Variants of the \ac{PF} algorithm have been used at the CELLO~\cite{Behrend:1982gk}, ALEPH~\cite{Buskulic:1994wz}, H1~\cite{H1:2020zpd}, ZEUS~\cite{Breitweg:1997aa,Breitweg:1998gc}, DELPHI~\cite{Abreu:1995uz}, CDF~\cite{Bocci:2001zx,Connolly:2003gb,Abulencia:2007iy}, D0~\cite{Abazov:2008ff}, CMS~\cite{Sirunyan:2017ulk,CMS:2008xjf} and ATLAS~\cite{Aaboud:2017aca} experiments to reconstruct a particle-level interpretation of high-multiplicity hadron collision events, given individual detector elements such as tracks and calorimeter clusters from a multi-layered, heterogeneous, irregular-geometry detector. 
The \ac{PF} algorithm generally associates tracks and calorimeter clusters from detector layers such as the \ac{ECAL}, \ac{HCAL} and others to reconstruct charged and neutral hadron candidates as well as photons, electrons, and muons with an optimized efficiency and resolution. 
Existing \ac{PF} reconstruction implementations are tuned using simulation for each specific experiment because detailed detector characteristics and geometry are critical for the best possible physics performance.

Recently, there has been significant interest in the use of supervised \ac{ML} to perform the  reconstruction in order to improve the physics reach of the experiments as well as reduce the computational requirements.
\ac{ML}-based reconstruction approaches using \acp{GNN}~\cite{gnn,gilmer2017neural,pointnet,Battaglia:2016jem,DGCNN} have been proposed for various tasks in particle physics~\cite{Shlomi:2020gdn,Farrell:2018cjr,Ju:2020xty,Amrouche:2019wmx,Amrouche:2019yxv,Choma:2020cry,Ju:2020tbo,Li:2020grn,Guo:2020vvt,Moreno:2019bmu,Moreno:2019neq,Qu:2019gqs,Mikuni:2020wpr,Qasim:2019otl,Martinez:2018fwc}, including \ac{PF} reconstruction~\cite{Kieseler:2020wcq,DiBello:2020bas,Duarte:2020ngm,Pata:2021oez}.
In particular in Ref.~\cite{Pata:2021oez}, a \ac{ML}-based \ac{PF} algorithm is proposed to reconstruct particle candidates in events with a large number of simultaneous \ac{PU} collisions using a simplified benchmark dataset generated with \PYTHIA~\cite{Sjostrand:2006za,Sjostrand:2007gs} and \DELPHES~\cite{deFavereau:2013fsa}.

\begin{figure}[htpb]
    \centering
    \includegraphics[width=0.8\linewidth]{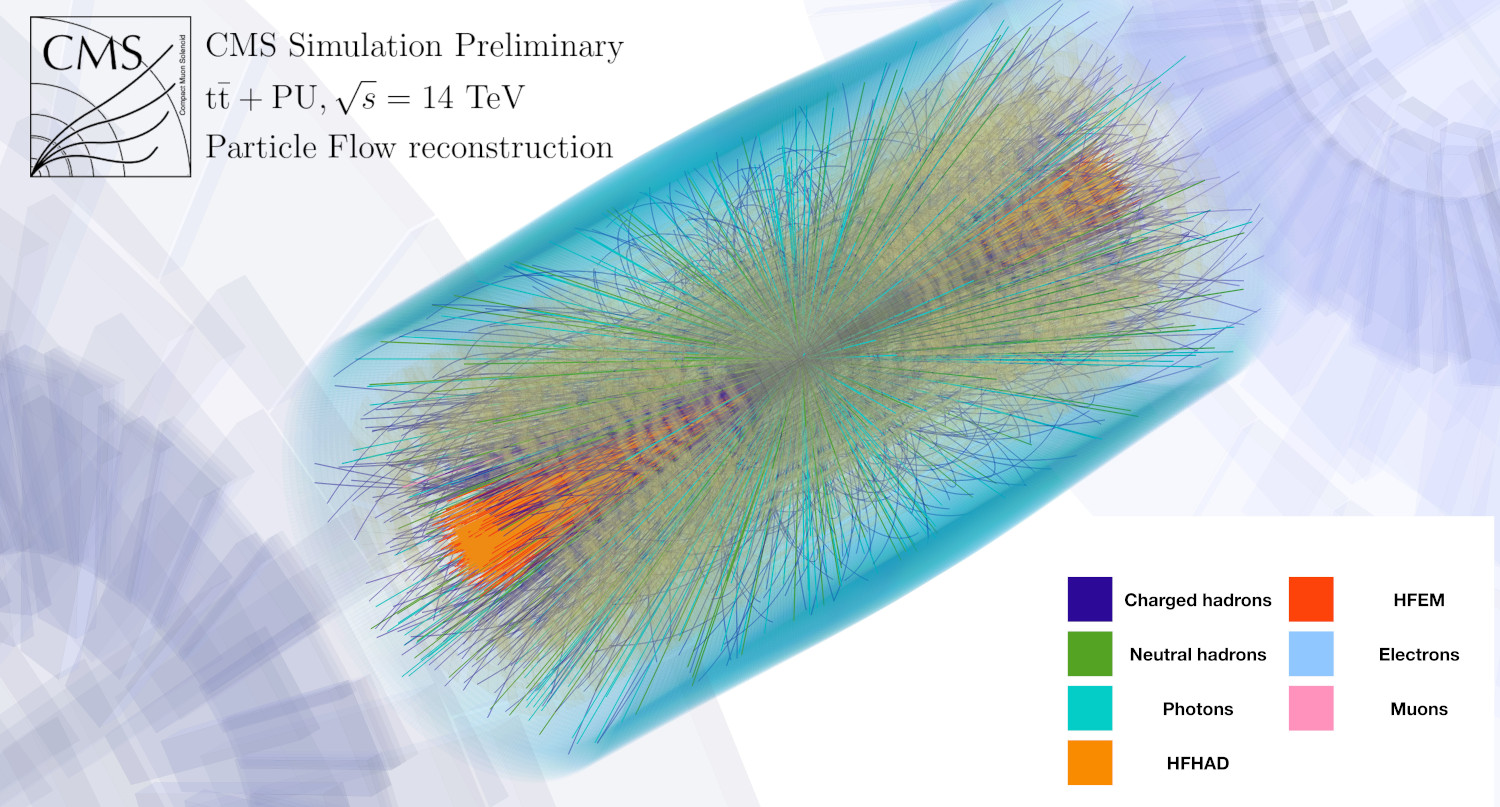}\\
    \includegraphics[width=0.8\linewidth]{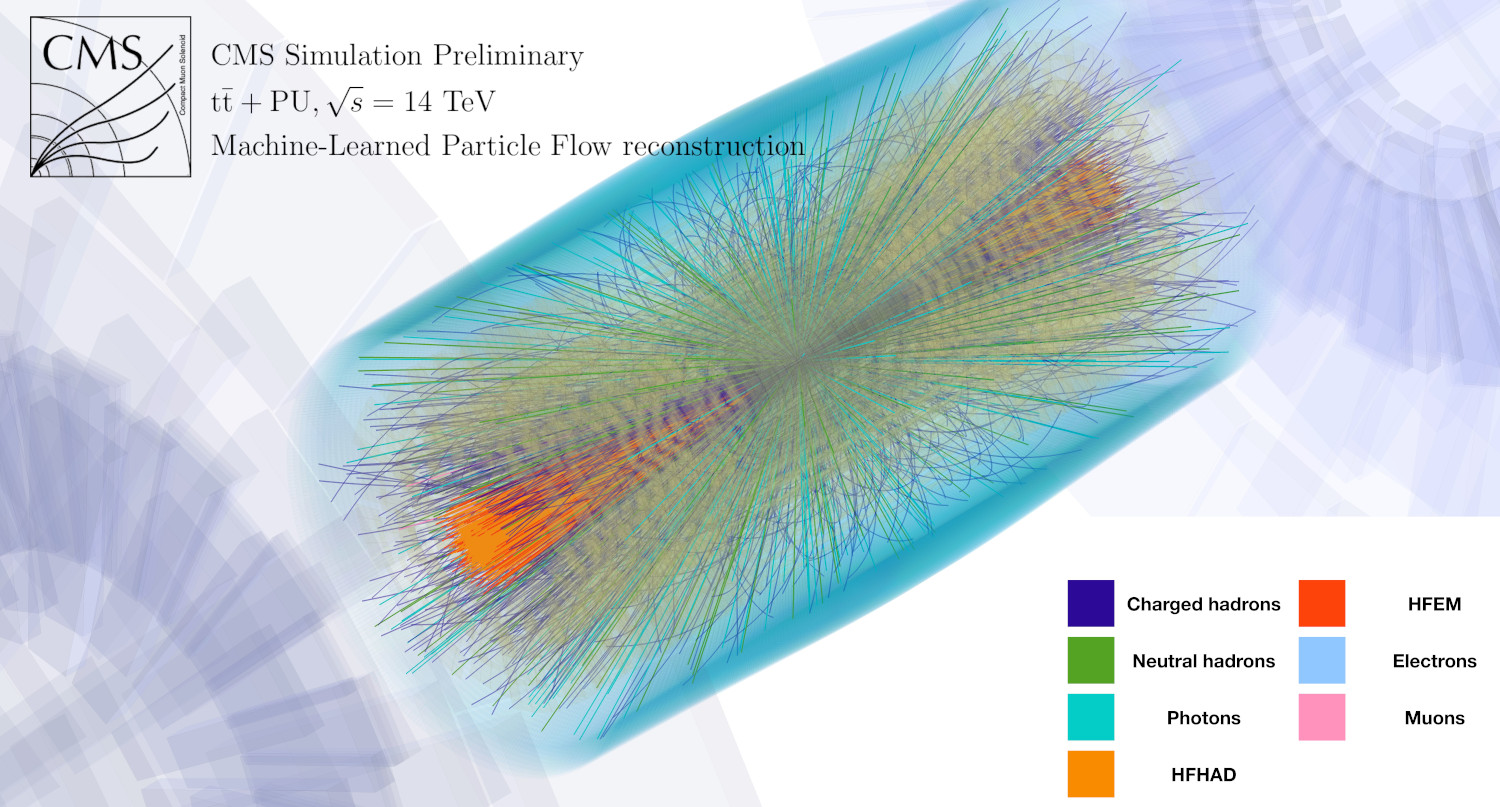}
    \caption{One simulated \ttbar event with pileup under \acs{LHC} Run 3 conditions, reconstructed with particle flow (top) and machine-learned particle flow (bottom). 
    The trajectories correspond to the particle flow candidates extrapolated to the ECAL surface, with candidates of different type shown in different colors. 
    We also show the ECAL detector surface (cyan) and the muon stations (blue).}
    \label{fig:one_event}
\end{figure}

In this work, we provide details on the implementation of a similar \ac{MLPF} algorithm for CMS~\cite{CMS-DP-2021-030}.
The CMS \ac{PF} algorithm reconstructs stable particles based on calorimeter clusters and tracks~\cite{CMS:2017yfk}.
An example reconstructed event, comparing \acs{PF} and \acs{MLPF}, is shown in~\cref{fig:one_event}. We present a possible evolution of \ac{PF} towards heterogeneous computing platforms such as GPUs using a \ac{GNN} in the following proceeding, structured as follows:
The training dataset is introduced in \cref{sec:dataset}.
We show the details of the \ac{MLPF} implementation and the computational performance of the algorithm in \cref{sec:model}.
In \cref{sec:results}, we validate the physics performance directly in the \ac{CMSSW} when the proposed algorithm is interfaced with the offline reconstruction of jets and missing transverse energy.
Finally, we provide a summary in \cref{sec:conclusions}.

\section{Dataset\label{sec:dataset}}
Simulated samples are used to train the \acs{MLPF} model.
For each event, we store the input detector signals and the target particles.
The input features for the detector signals are chosen based on the \ac{PF} reconstruction algorithm~\cite{CMS:2017yfk}:
\begin{itemize}
    \item \ac{ECAL}, \ac{HCAL}, \ac{HF} calorimeter clusters: cluster energy, corrected energy, $\eta$, $\phi$, $x$, $y$, $z$ position; number of hits; layer; depth; cluster flags
    \item \ac{ECAL} supercluster: cluster energy, $\eta$, $\phi$, $x$, $y$, $z$ position; number of hits
    \item \ac{KF} tracks: \ptmomentum, $\eta$, $\phi$, $p_x$, $p_y$, $p_z$, $|p|$ at the vertex; $\eta$, $\phi$ extrapolated to the \ac{ECAL} shower max and \ac{HCAL} entrance; number of hits; track charge; number of \ac{DT}, \ac{CSC} hits
    \item \ac{GSF} tracks: \ptmomentum, $\eta$, $\phi$, $p_x$, $p_y$, $p_z$, $E$ at the inner point; $\eta$, $\phi$ at the outer point; track charge; number of hits; a flag to denote if the electron seed is \ac{ECAL} or tracker driven.
    \item \ac{BREM} points: index of the trajectory point, $\Delta P$, $\sigma(\Delta P)$
\end{itemize}
We choose the existing \ac{PF} particles as the regression target in order to constrain the problem to a well-defined and well-understood detector response.
It is clear that the \ac{MLPF} model trained on the baseline \ac{PF} as the target does not allow to exceed \ac{PF} performance, but we expect to broadly reproduce the existing physics response both on particle level as well as for jets and \ac{MET}.
In addition, this allows to test the computational performance and integration with downstream reconstruction.
Training on a generator-level set of target particles is left for a future study.

We use a mixture of simulated physics collisions such as \ttbar with \ac{PU}, as well as simulations of single particles shot from the interaction point.
All samples are generated under identical Run 3 conditions.
We split the simulated samples into a training (80\%) and test set (20\%) after randomizing the order.
The sample sizes are reported in~\cref{tab:samples}.

\begin{table}[ht]
    \centering
    \caption{Simulation samples used for optimizing the \acs{MLPF} model.}
    \begin{tabular}{c|c|c}
        Sample fragment & \acs{PU} Configuration & \acs{MC} events \\
        \hline
        Top quark-antiquark pairs (\ttbar) & Flat 55--75 & 20\,k \\
        $\PZ\to\tau\tau$ all-hadronic & Flat 55--75 & 20\,k \\
        Single electron flat $\ptmomentum\in[1,100]\GeV$ & No \acs{PU} & 400\,k \\
        Single muon flat $\ptmomentum\in[0.7,10]\GeV$ & No \acs{PU} & 400\,k \\
        Single $\pi^0$ flat $\ptmomentum\in[0,10]\GeV$ & No \acs{PU} & 400\,k \\
        Single $\pi$ flat $\ptmomentum\in[0.7,10]\GeV$ & No \acs{PU} & 400\,k \\
        Single $\tau$ flat $\ptmomentum\in[2,150]\GeV$ & No \acs{PU} & 400\,k \\
        Single $\gamma$ flat $\ptmomentum\in[10,100]\GeV$ & No \acs{PU} & 400\,k
    \end{tabular}
    \label{tab:samples}
\end{table}

\section{Model and Optimization\label{sec:model}}

The task in the \acs{MLPF} setup is to predict the set of particles $y'_k\in Y'$ in the event, given the set of detector signals $x_i \in X$.
The inputs comprise the features of \acs{ECAL}, \acs{HCAL}, \acs{HF} calorimeter clusters, \acs{ECAL} superclusters, \acs{KF} tracks, \acs{GSF} tracks, and \acs{BREM} points described in \cref{sec:dataset}.
The target particles $y_i$ are described by a feature vector $y_k = [\mathrm{ID}, \ptmomentum, \eta, \phi, E, q]$, where $\mathrm{ID}$ is a one-hot encoded vector representing the \acs{PF} particle candidate type among eight options: charged hadron, neutral hadron, \ac{HFEM}, \ac{HFHAD}, photon, electron, muon, or none.

The model is optimized with respect to the set of true target particles $y_k\in Y$ in each event.
In order to practically compute the loss function between two sets of arbitrary (and possibly different) size, we follow the object condensation approach first introduced in Ref.~\cite{Kieseler:2020wcq}, and as implemented for particle flow reconstruction in Ref.~\cite{Pata:2021oez}.
Effectively, we first zero-pad the target set $Y$ such that $|Y|=|X|$ and evaluate how well each predicted particle's type is classified and momentum is regressed.
This per-particle loss is a physics-based simplification of the generic set-to-set loss function.
The total loss function for one event is then $L(Y,Y') = \sum_{k=1}^{|X|} \mathrm{focal}(y_k, y'_k) + \mathrm{Huber}(y_k, y'_k)$, where we use the focal~\cite{DBLP:journals/corr/abs-1708-02002} and Huber~\cite{10.1214/aoms/1177703732} losses for classification and momentum regression, respectively.

The \acs{MLPF} \ac{GNN} model is implemented in \TENSORFLOW~\cite{tensorflow2015-whitepaper} (the latest stable version, v2.6, at the time of writing) and can be exported to \ONNX~\cite{onnx} for inference. 
Only standard matrix operations using dense arrays are used, thus it is expected to be highly portable across platforms.
Particular care is taken to ensure scalability of the algorithm by using local context binning.

% \begin{figure}[htpb]
%     \centering
%     \includegraphics[width=0.4\linewidth]{figures/bins_cg_0.pdf}
%     \includegraphics[width=0.4\linewidth]{figures/bins_cg_1.pdf}
%     \caption{The learned binning structure in the first two layers of the model.
%     We show one simulated \ttbar event, with each point corresponding to a PFElement in the event.
%     The colors correspond to the assignment of the PFElements into the bins in each layer.}
%     \label{fig:model_bins}
% \end{figure}

\begin{figure}[htbp]
    \centering
    \includegraphics[width=0.8\linewidth]{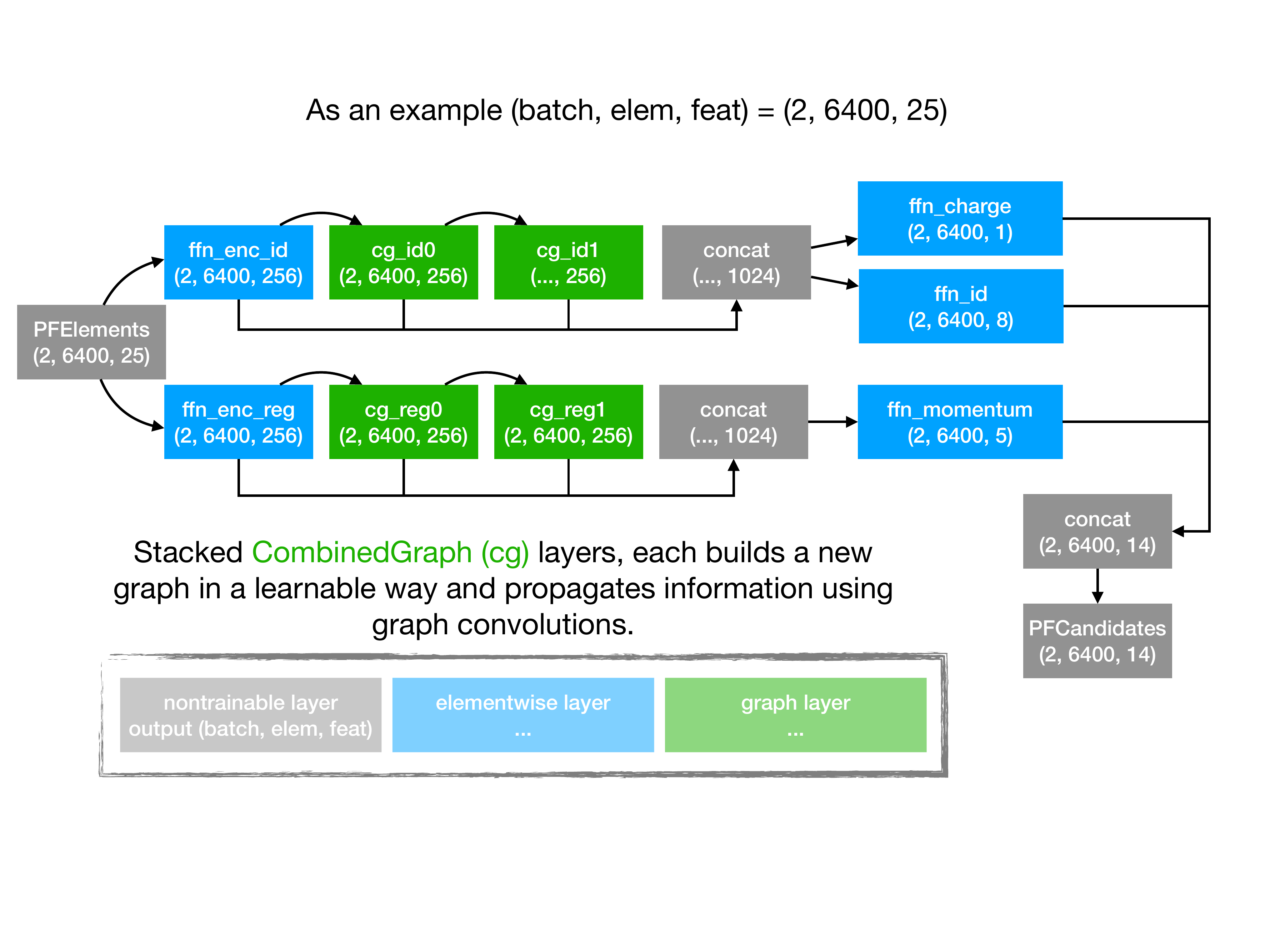}\\
    \includegraphics[width=0.8\linewidth]{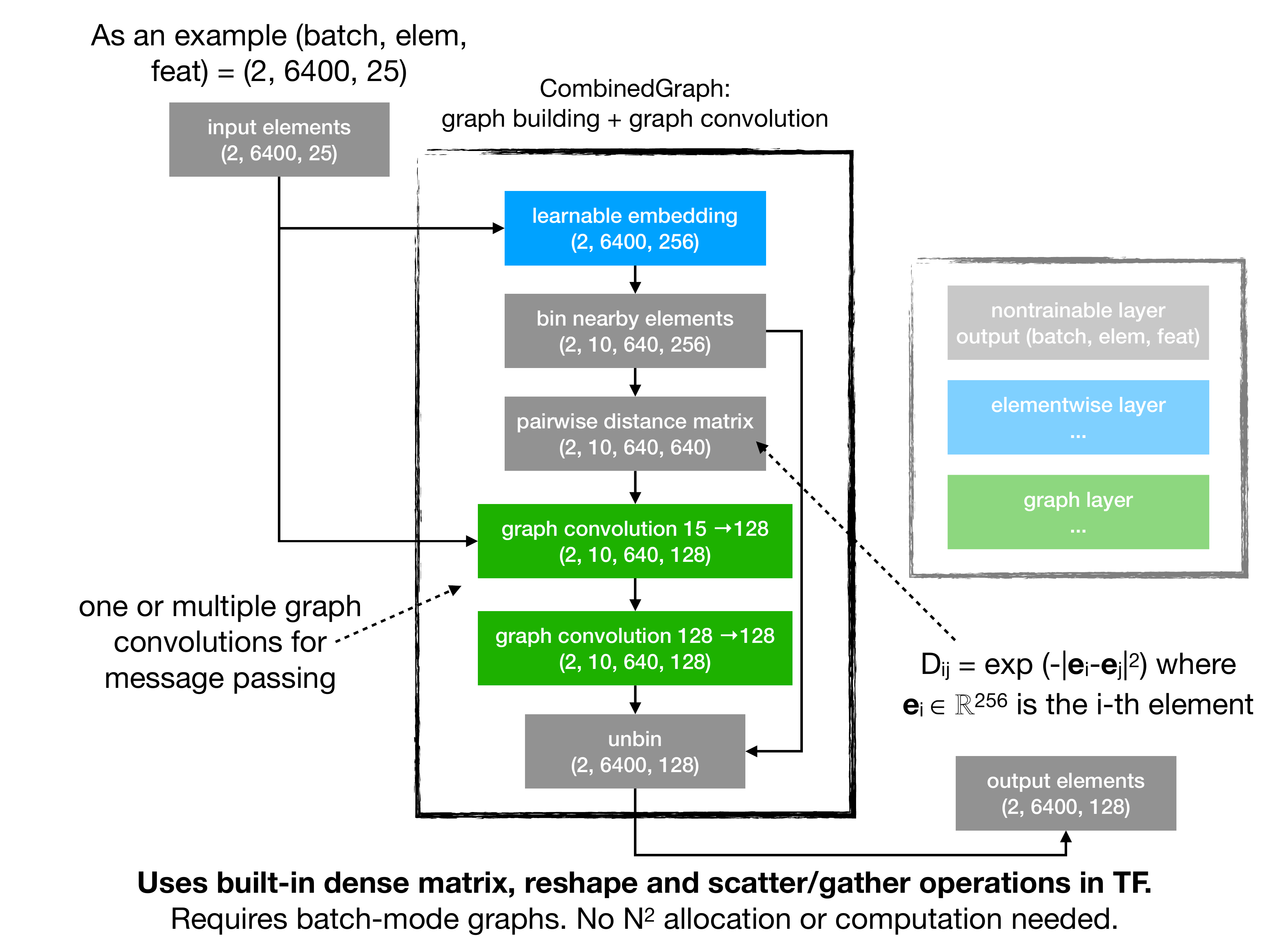}
    \caption{A schematic overview of the scalable, multi-layered graph neural network as implemented in \TENSORFLOW (top) and the scalable combined graph layer (bottom).
    The model consists of a classification and regression branch. 
    Both branches consist of multiple scalable graph building and convolution layers. 
    Elementwise feedforward networks are used to decode the encoded inputs to a classification and regression prediction.
    The input elements are projected into a learnable embedding space. 
    Nearby elements in the embedding space are binned to fixed-size bins. 
    A fully-connected graph is built in each bin, which is used for one or multiple graph convolutions that are used to transform the input elements.
    Finally, the transformed elements are unbinned.}
    \label{fig:model_structure}
\end{figure}

The local context binning is inspired by the existing \acs{PF} block algorithm, but has been reformulated to be optimizable using \acs{ML} and ensures the approximately linear scaling of the runtime and memory with the input size.
In each local context bin, a graph is built dynamically using a Gaussian kernel.
Once the graphs are built dynamically in the event, information can be propagated between the elements in a learnable fashion.
For generality, several layers of graph building and message propagation can be stacked.
A schematic overview of the network architecture and the scalable combined graph layer is presented in~\cref{fig:model_structure}.
The code to build, train, and evaluate the model is publicly available~\cite{joosep_pata_2021_5520559}.

The best performing \acs{MLPF} hyperparameters were found after two separate hyperparameter searches.
The first search used the Bayesian optimization hyperband (BOHB) algorithm~\cite{DBLP:journals/corr/abs-1807-01774} to tune parameters: the learning rate, learning rate schedule, and dropout percentage. 
The second search fixed the best hyperparameters found in the first search and used the asynchronous successive halving algorithm (ASHA) \cite{DBLP:journals/corr/abs-1810-05934} to tune model-architecture related parameters: the number of graph layers, the sizes of input encoding and output decoding layers and the linearization bin size.
ASHA allows for an efficient use of compute resources when performing distributed multi-worker hypertuning by aggressively early stopping trials that underperform relative to other trials.
More details about the hyperparameter optimization can be found in~\cite{opt-proceedings}.

\section{Results\label{sec:results}}
The \acs{MLPF} model is interfaced with offline reconstruction in \acs{CMSSW}, though only the standard \ac{PF} reconstruction is used for CMS reconstruction by default.
For inference, the model is exported from \TENSORFLOW to \ONNX, such that the native \ONNXRUNTIME that is already integrated with \acs{CMSSW} can be used.
The validation plots that follow are independent of the training setup as well as the training samples.
This ensures that the model is tested under realistic conditions in actual physics reconstruction.
We validate the model on both \ttbar and \qcd events, the latter in particular is to verify the generalization capabilities of the \acs{ML}-based approach for particle reconstruction, as \qcd events have a different momentum distribution compared to \ttbar and were never seen during training.

\begin{figure}[htpb]
    \centering
    \includegraphics[width=0.45\linewidth]{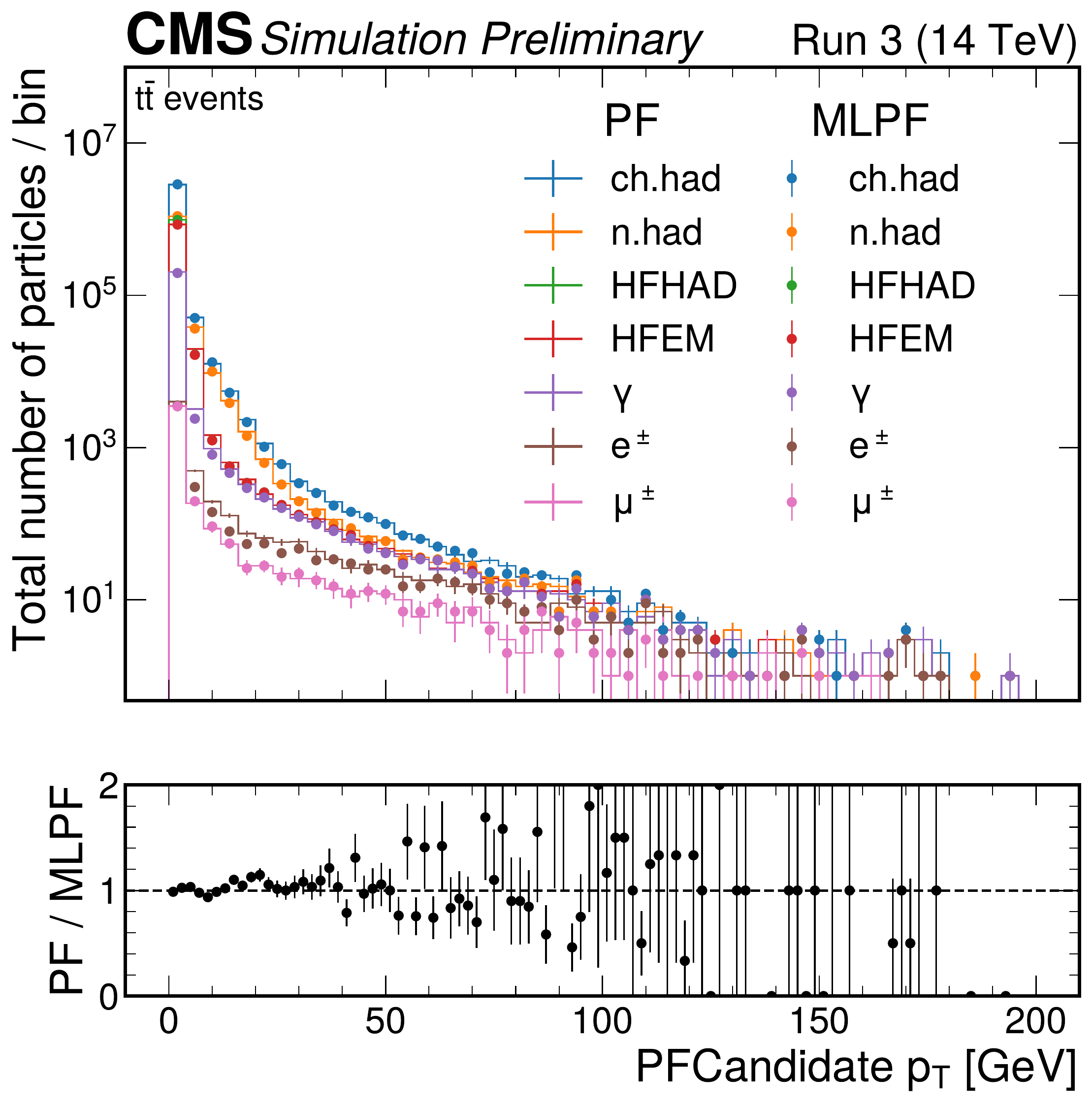}
    \includegraphics[width=0.45\linewidth]{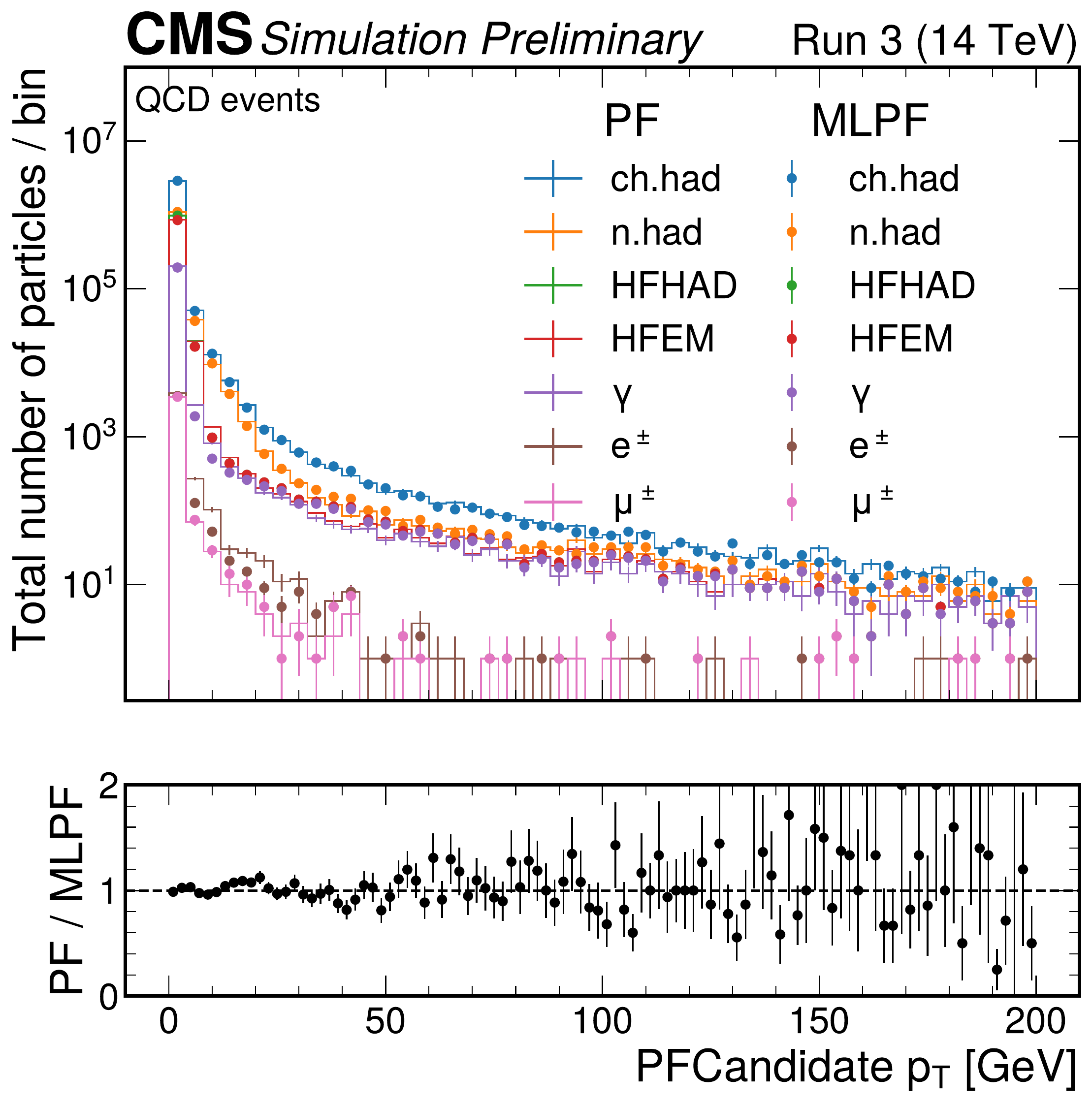}\\
    \includegraphics[width=0.45\linewidth]{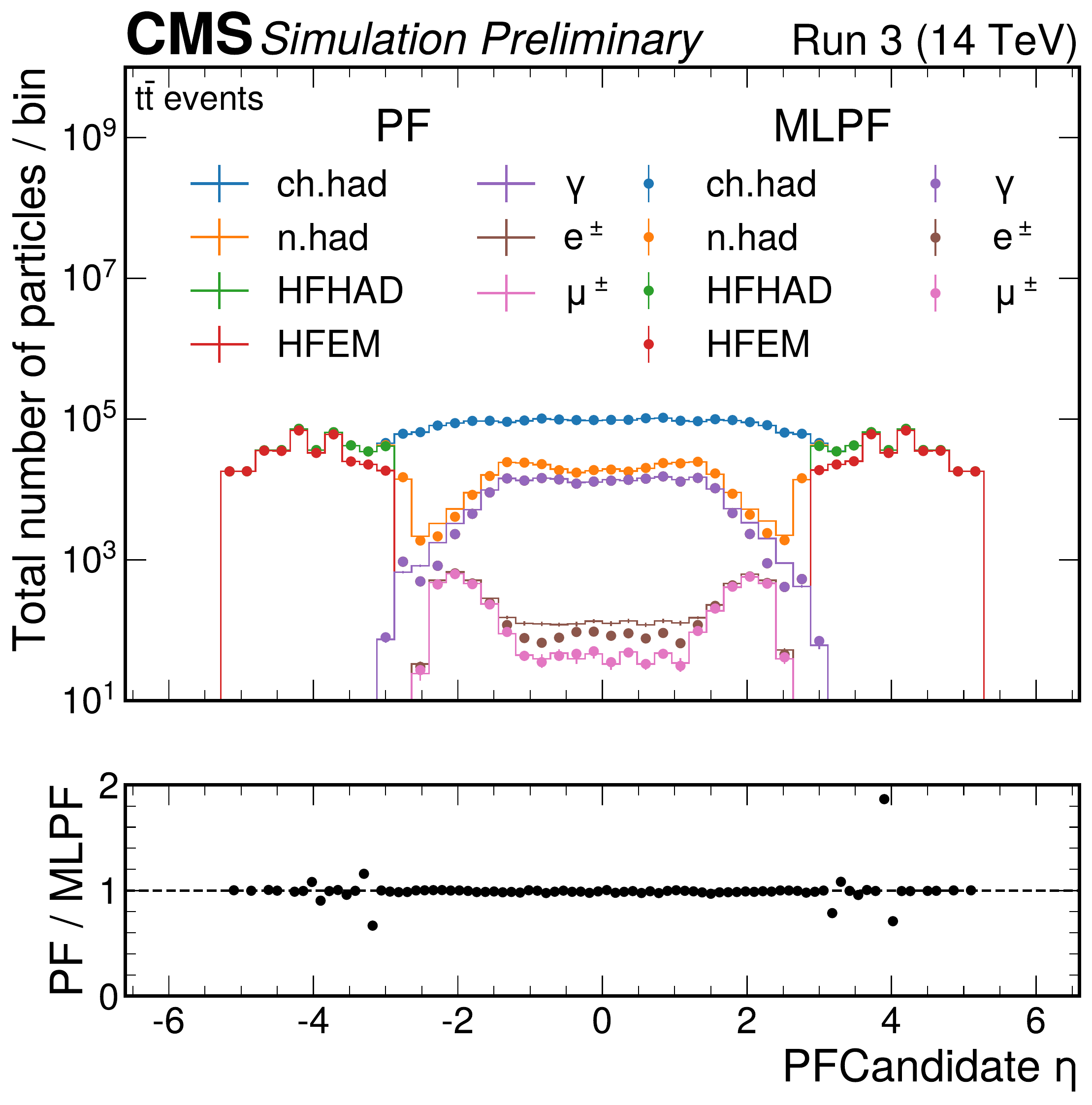}
    \includegraphics[width=0.45\linewidth]{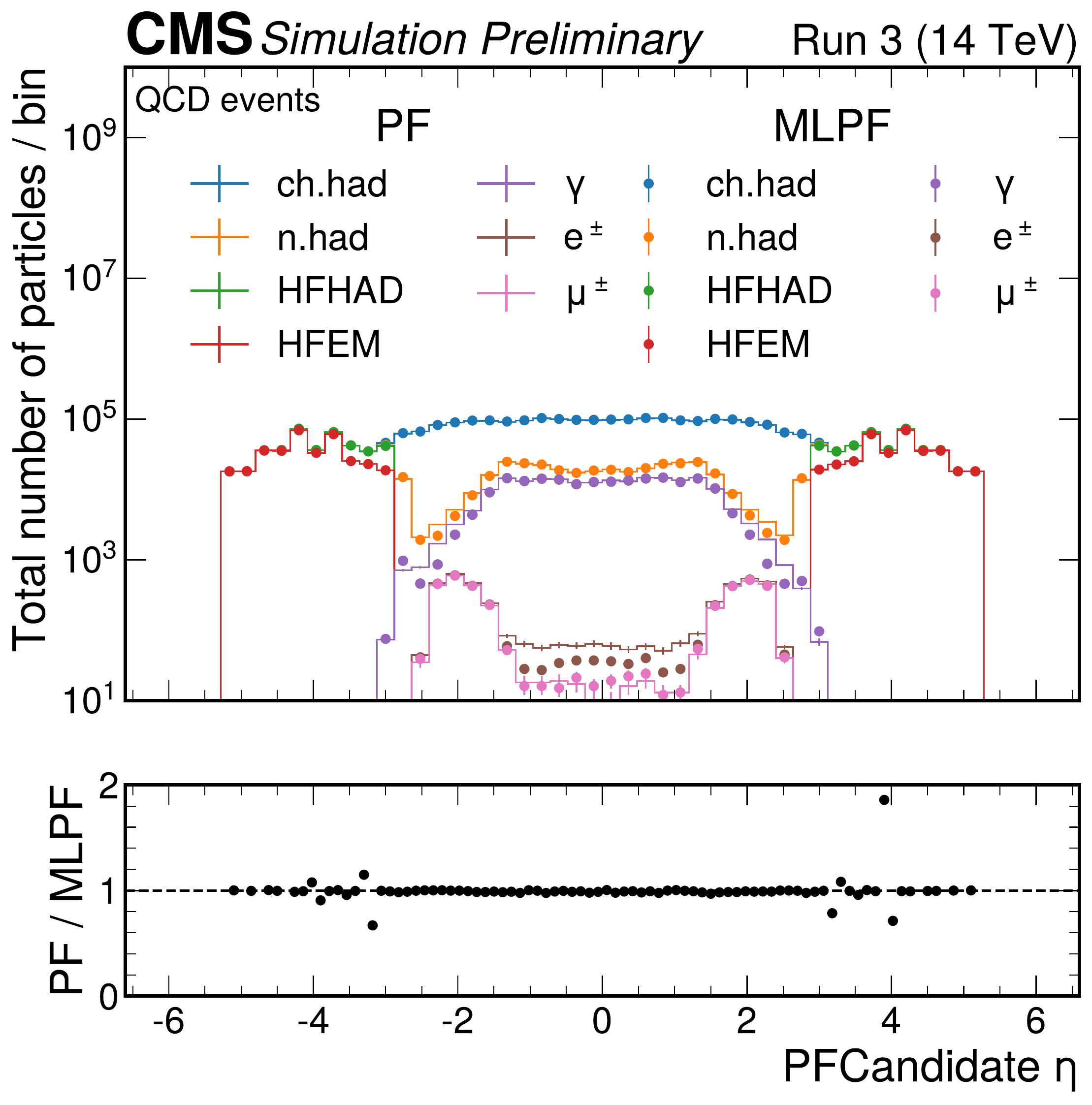}\\
    \caption{Reconstructed particle candidate \ptmomentum (top), \pseudorapidity (bottom) distributions between PF (lines) and MLPF (dots) for \ttbar (left) and \qcd (right). Different reconstructed particle candidate types are shown using colors.}
    \label{fig:cmssw_candidates}
\end{figure}

Particle-level comparisons between \acs{PF} and \acs{MLPF} candidates are shown in~\cref{fig:cmssw_candidates}.
In general, we observe a good correspondence between the \ptmomentum and \pseudorapidity distributions for all particle types, within the available statistics, except neutral hadrons in the $\ptmomentum < 10\GeV$ regime, and the overall electron multiplicity.

\begin{figure}[htpb]
    \centering
    \includegraphics[width=0.45\linewidth]{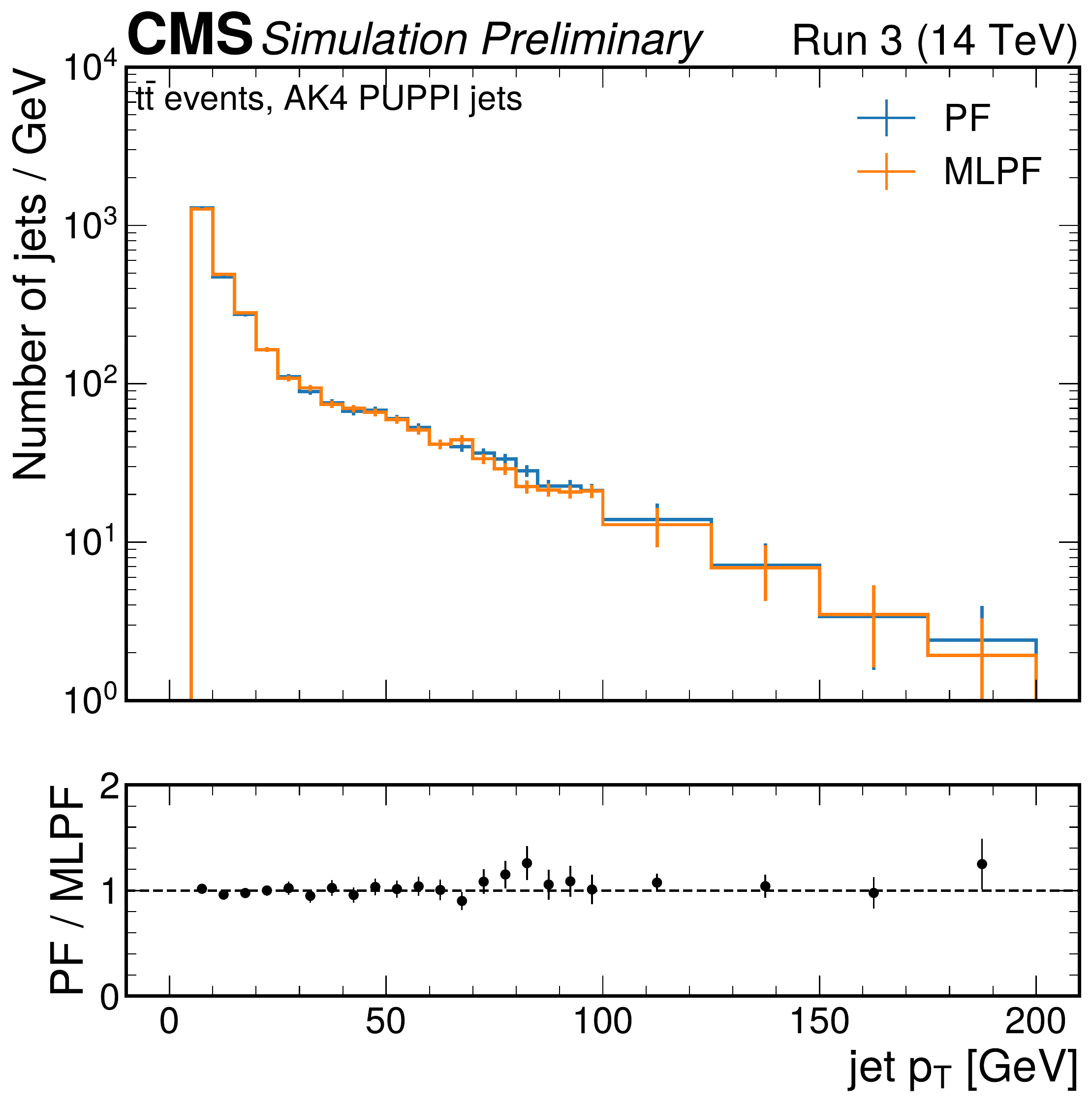}
    \includegraphics[width=0.45\linewidth]{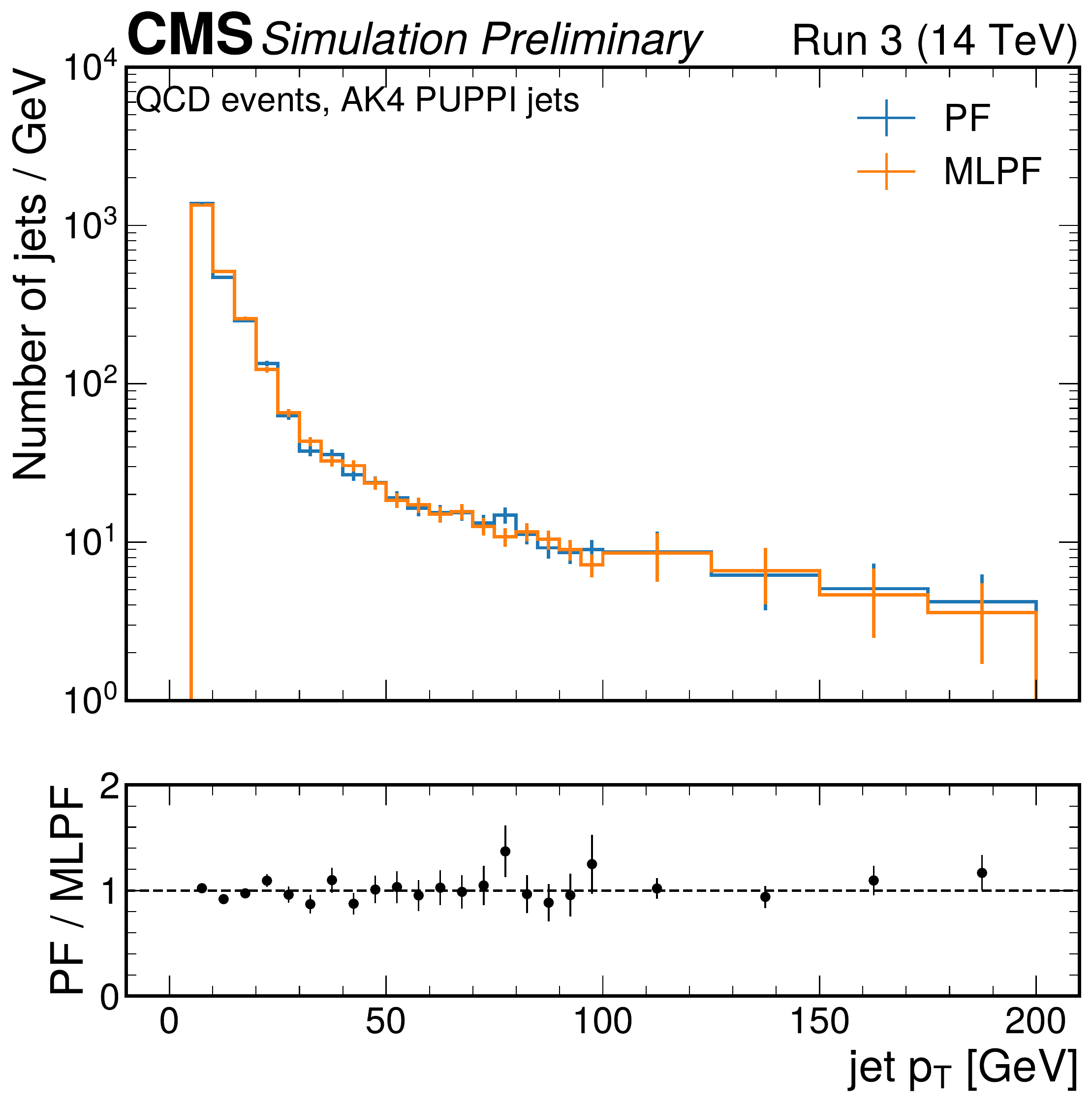}\\
    \includegraphics[width=0.45\linewidth]{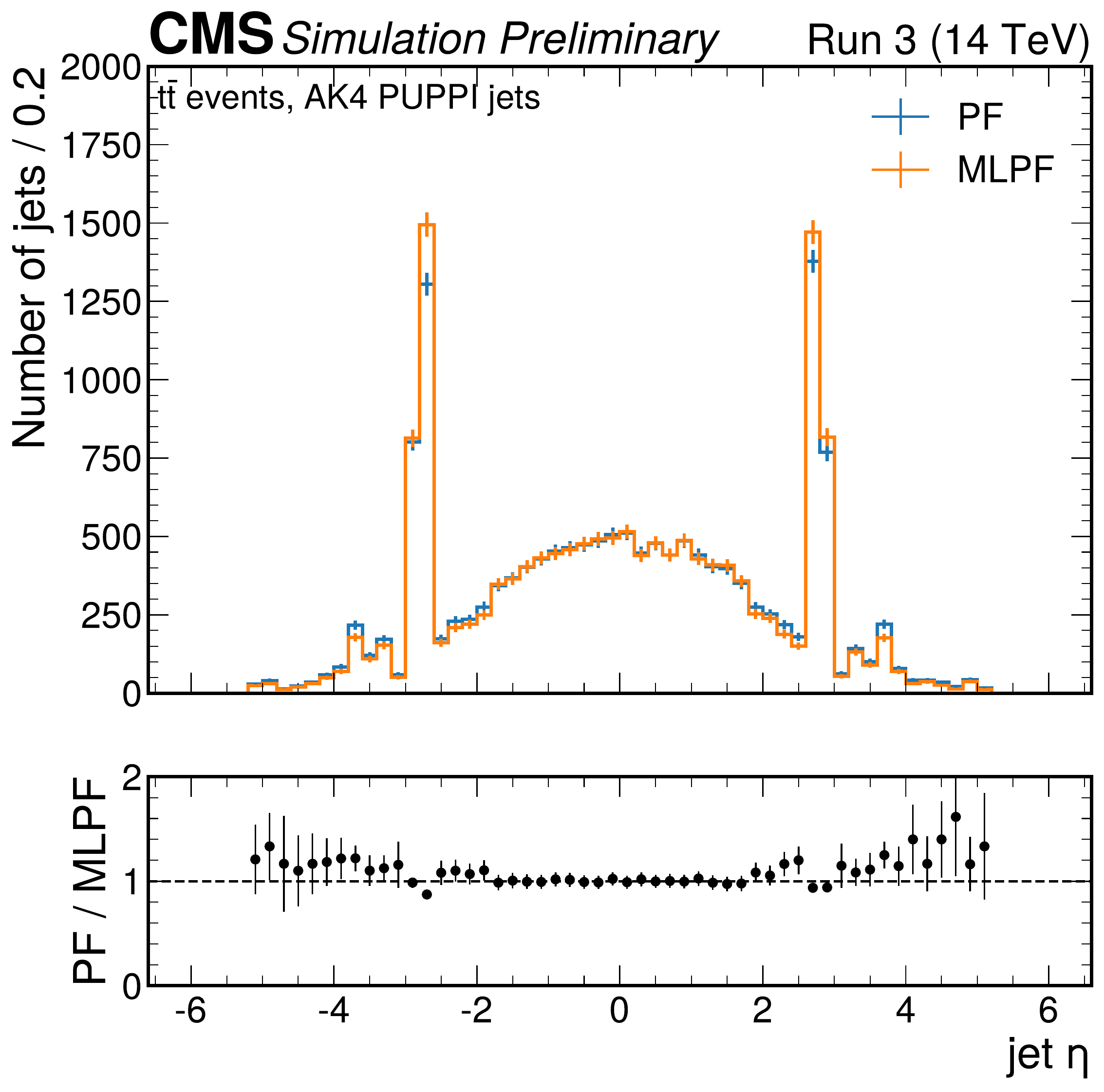}
    \includegraphics[width=0.45\linewidth]{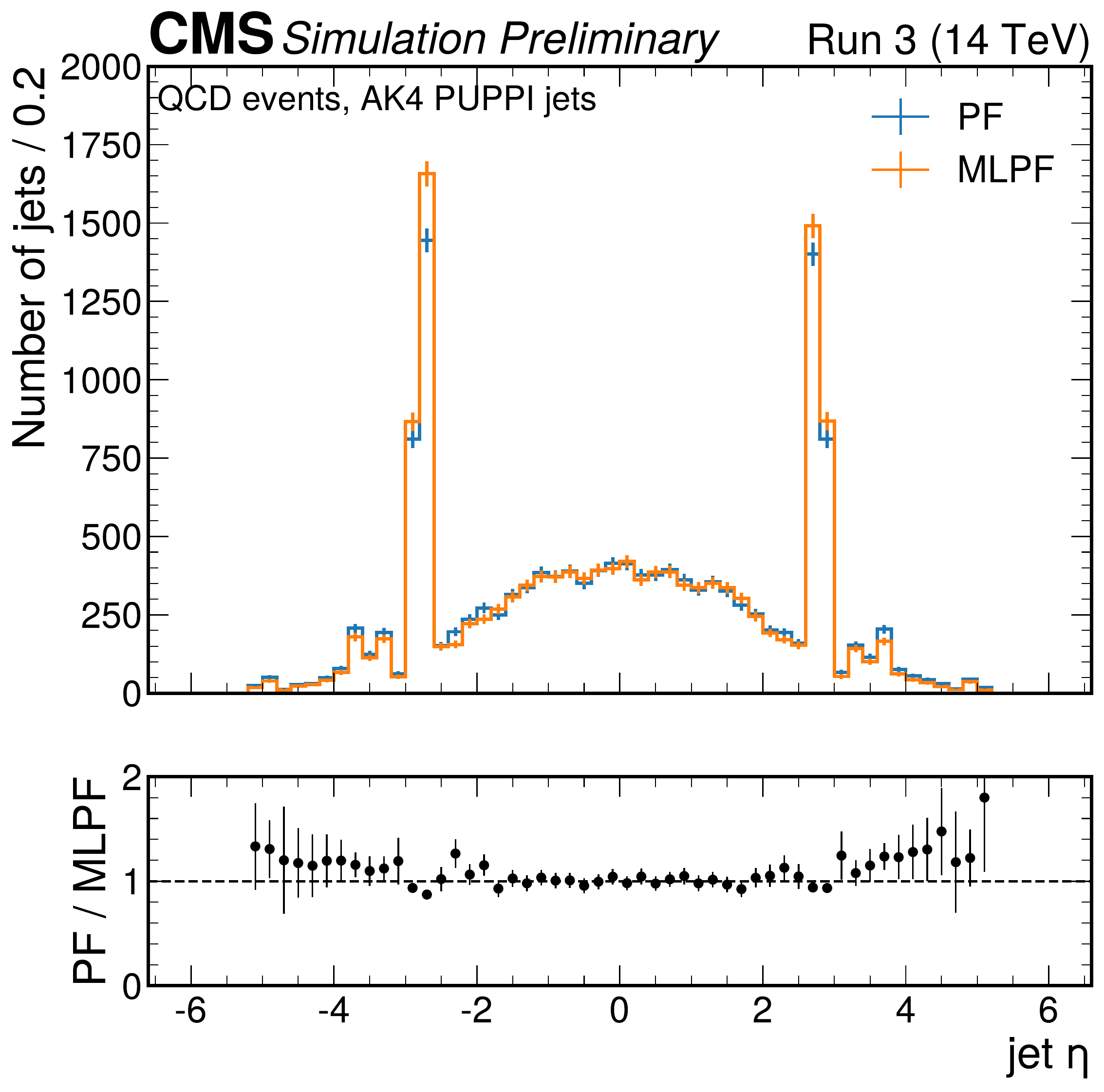}\\
    \caption{Transverse momentum distributions for \akfourpuppi~\cite{Cacciari:2005hq,Bertolini:2014bba} jets, with PF and MLPF, for \ttbar (left column) and \qcd (right column).}
    \label{fig:dqm_jet}
\end{figure}

\begin{figure}[htpb]
    \centering
    \includegraphics[width=0.45\linewidth]{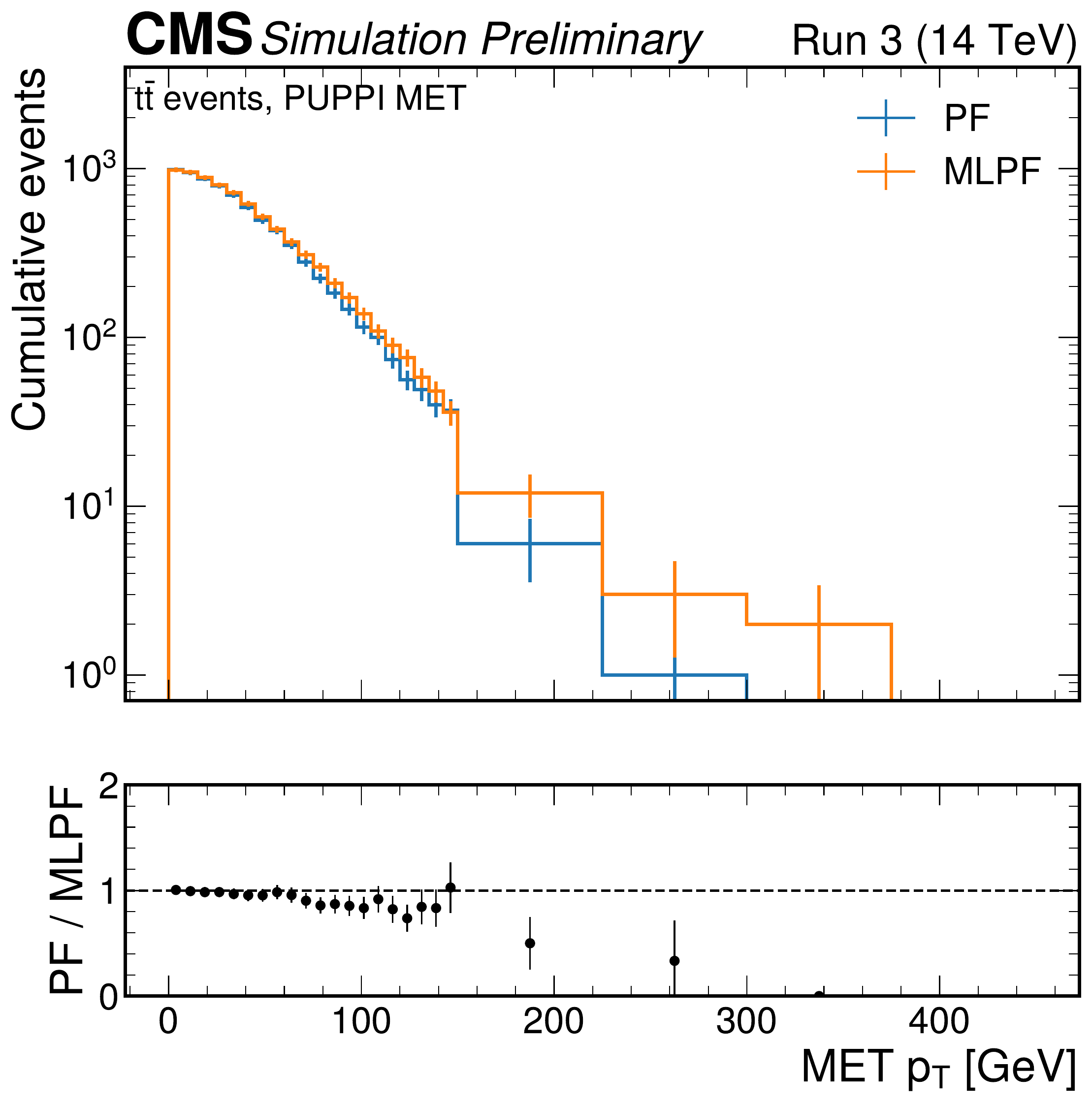}
    \includegraphics[width=0.45\linewidth]{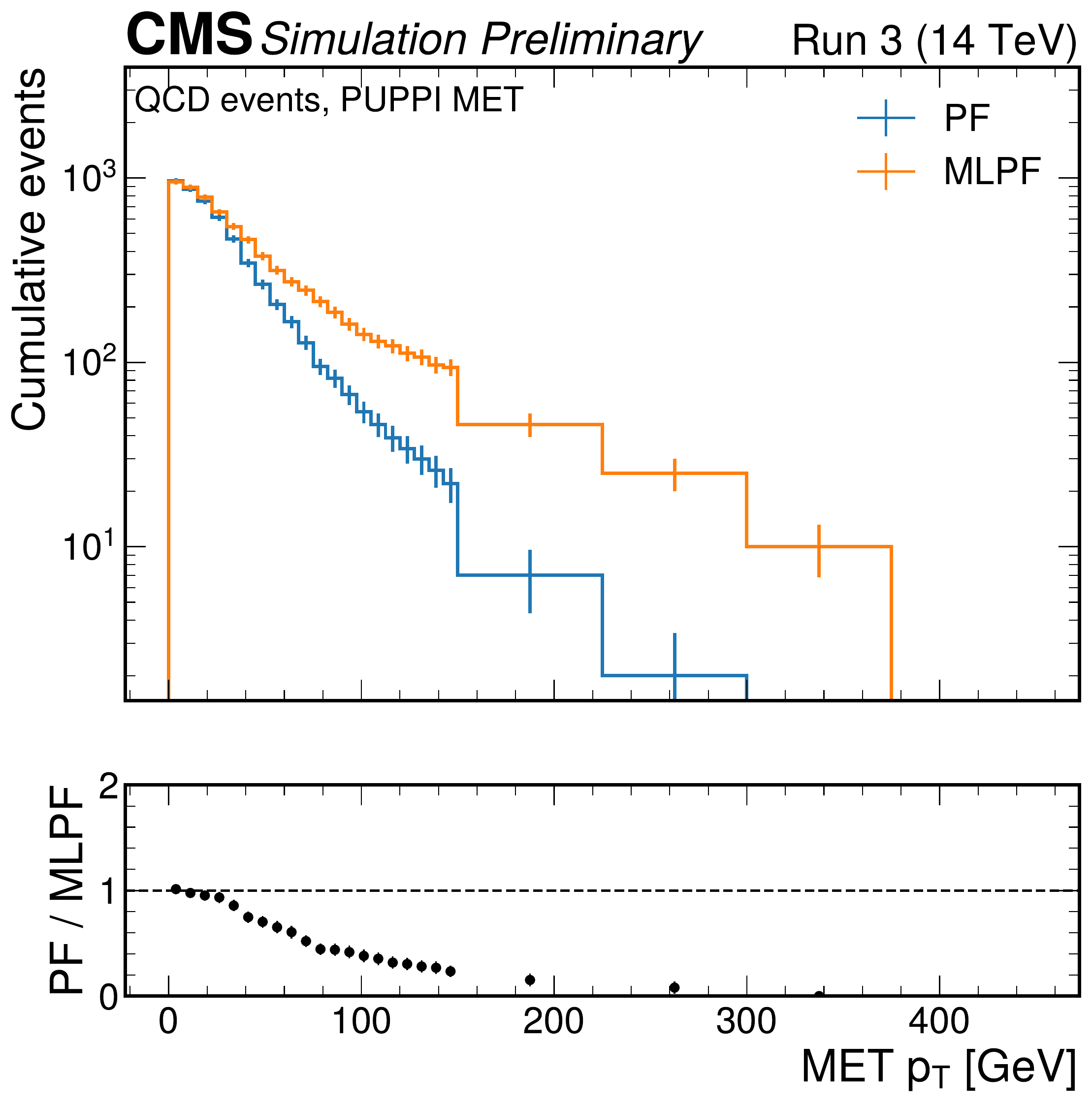}\\
    \caption{Missing transverse energy distribution, with PF and \acs{MLPF}, for \ttbar (left) and \qcd (right).}
    \label{fig:dqm_met}
\end{figure}

Object-level comparisons (jets, \acs{MET}) are shown in~\cref{fig:dqm_jet,fig:dqm_met}. 
In general, we observe a good correspondence between the baseline \acs{PF} and the proposed \acs{MLPF} algorithm in the bulk of the distributions, within the available statistics.
However, for \acs{MET}, we observe a misreconstructed high-\acs{MET} tail that is most prominent for the \qcd sample not used in training.
This could potentially be attributed to limited training statistics for high-energy neutral particles, which are correspondingly not well reconstructed by the \acs{MLPF} algorithm in the current iteration and require further study.

\begin{figure}[htpb]
    \centering
    \includegraphics[width=0.8\linewidth]{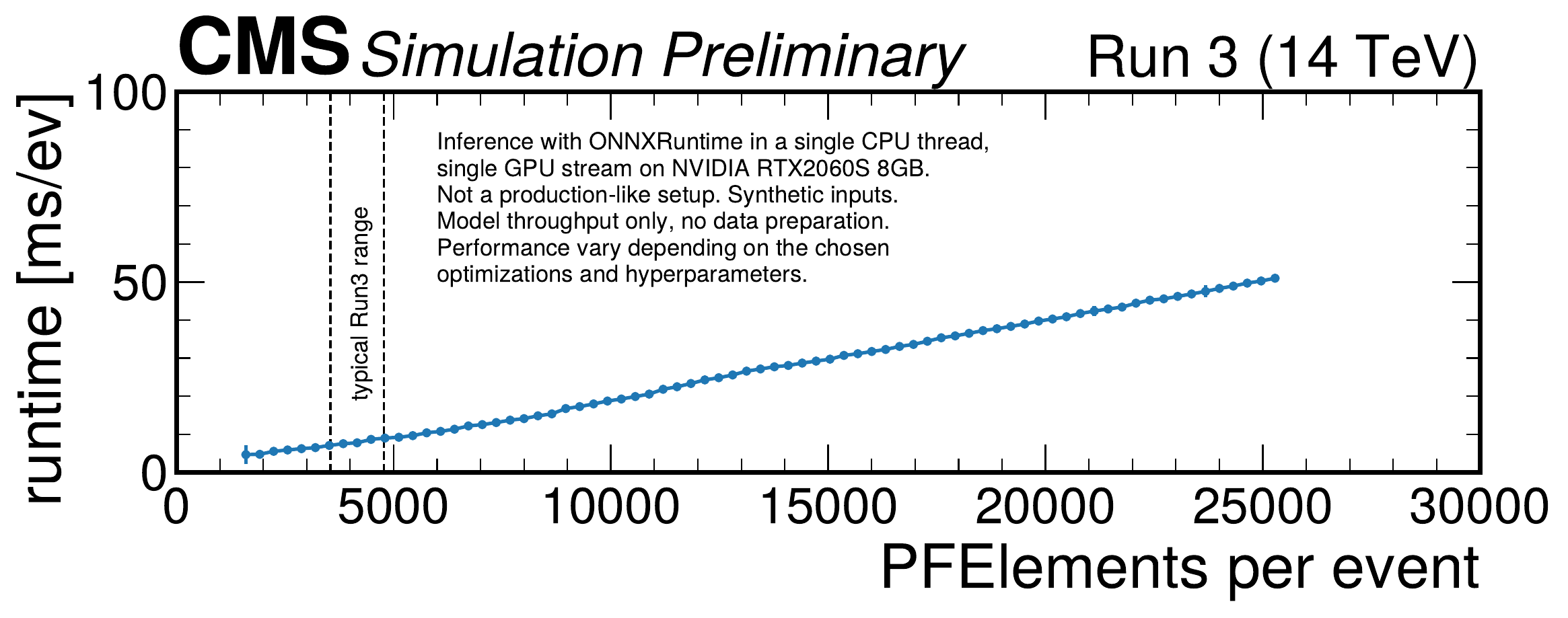}\\
    \includegraphics[width=0.8\linewidth]{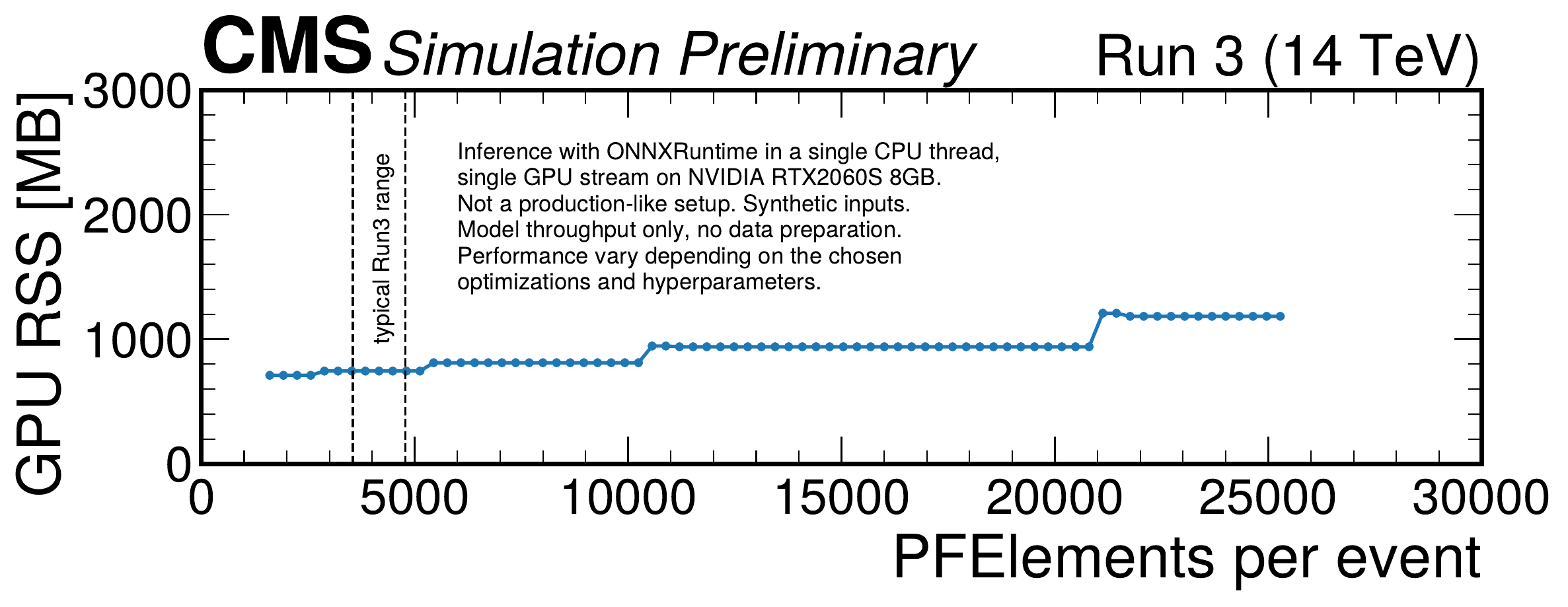}\\
    \caption{Runtime and memory use on an NVIDIA RTX2060S GPU with a varying event size. 
    The $1\sigma$ range of a typical Run3 \ttbar event with \acs{PU} is shown with the dashed lines. 
    Note that this is not a production-like setup, as a single GPU execution stream is used at a time. 
    A realistic production-like test would involve the GPU running multiple different models/kernels from multiple execution streams in parallel.}
    \label{fig:computational_scaling}
\end{figure}

We report the computational performance of the model on a single stream on a single GPU in~\cref{fig:computational_scaling}.
We observe an approximately linear dependence of runtime and memory consumption with increasing particle multiplicity, with a typical Run 3 event requiring around 10\unit{ms} of wall time and around 1\unit{GB} of RAM on a GPU\footnote{Due to the still-evolving support for \acs{GPU} \ac{ML} evaluation in \ac{CMSSW}, this measurement was carried out in a standalone environment, outside the standard reconstruction sofware.}.

\section{Summary\label{sec:conclusions}}
\acresetall
We have developed a \acs{GPU}-native algorithm for \acs{PF} reconstruction at CMS and presented a first integration with offline reconstruction.
The new \acs{MLPF} algorithm is based on a supervised \acs{ML} setup, where we optimize a scalable \acs{GNN} model to reconstruct the output particles of the current \acs{PF} algorithm, based on the input detector elements.
In general, we observe a high degree of correspondence between the \acs{MLPF} algorithm and the baseline \acs{PF} both at the particle level, as well as at the object level in jets and \acs{MET}, when the new model is interfaced with offline reconstruction in \acs{CMSSW}.
Some differences are observed in neutral hadron and electron performance, which we plan to address with additional training statistics and optimization.
The \acs{MLPF} model has an approximately linear scaling of runtime and memory with increasing particle multiplicity.

\ack
We thank our colleagues in the CMS Collaboration, especially in the Particle Flow, Physics Performance and Dataset, Offline and Computing, and Machine Learning groups, in particular Kenichi Hatakeyama, Lindsey Gray, Jan Kieseler, Danilo Piparo, Gregor Kasieczka, Salvatore Rappoccio, Kaori Maeshima, Kenneth Long, and Juska Pekkanen for helpful feedback in the course of this work.
JP was supported by the Mobilitas Pluss Grant No. MOBTP187 of the Estonian Research Council.
JD and FM were supported by DOE Award Nos. DE-SC0021187 and DE-SC0021396 (FAIR4HEP).
FM was also supported by a UCSD HDSI fellowship and an IRIS-HEP fellowship through NSF Cooperative Agreement OAC-1836650. EW was supported by the CoE RAISE Project which have received funding from the European Union’s Horizon 2020 – Research and Innovation Framework Programme H2020-INFRAEDI-2019-1 under grant agreement no. 951733.

\clearpage
\printbibliography

\end{document}